\documentclass[a4paper,12pt]{article}
\usepackage{graphicx}
\usepackage[T1,T2A]{fontenc}
\usepackage[utf8]{inputenc}
\usepackage{amssymb}
\usepackage{amsmath}
\usepackage{setspace}
\usepackage{xcolor}
\graphicspath{{images/}}
\begin{document}

\thispagestyle{empty}

\begin{center}
{\large
{\bf 
Critical behavior of a 2D spin-pseudospin model in a strong exchange limit
} 
}
\vskip0.5\baselineskip{
\bf 
D.~N.~Yasinskaya$^{*1}$, 
V.~A.~Ulitko$^{1}$,
A.~A.~Chikov$^{1}$,
Yu.~D.~Panov$^{1}$
}
\vskip0.1\baselineskip{
$^{1}$Ural Federal University 620002, 19 Mira Street,  Ekaterinburg, Russia
}
\vskip0.1\baselineskip{
$^{*}$daria.iasinskaia@urfu.ru
}
\end{center}

We study the 2D static spin-pseudospin model equivalent to the dilute frustrated antiferromagnetic Ising
model with charge impurities. We present the results of classical Monte Carlo simulation on a square lattice
with periodic boundary conditions in a ``strong'' exchange limit. In the framework of the finite-size scaling theory we obtained the static critical exponents for the specific heat $\alpha$  and the correlation length $\nu$ for a wide range of the local density–density interaction parameter $\Delta$ and charge density $n$. It was shown that the system exhibits non-universal critical behavior depending on these parameters.\\


\section{Introduction}
Magnetic materials with competitive interactions or geometric frustration are being investigated for past several decades because of a rich variety of exotic magnetic phases and unconventional critical behavior~\cite{Starykh}. A frustration may apply different phase-transition scenarios, nonuniversality, new critical points~\cite{Honecker}. Additionally, the effect of impurities may lead to a crossover to a novel critical behavior~\cite{Dotsenko}.  Coexistence and competition of charge and spin orderings is typical for high-T$_c$ cuprates~\cite{topical}. To study the spin-charge competition in cuprates such as La$_{2-x}$Sr$_x$CuO$_4$, a simple static spin-pseudospin model has recently been proposed~\cite{Mod1,Mod2}. This model is equivalent to the frustrated antiferromagnetic 2D Ising model with charged impurities. The model considers the CuO$_2$ planes as a system of charge triplets Cu$^{1+,2+,3+}$, associated with the three projections of the $S=1$ pseudospin. At variance with nonmagnetic hole (Cu$^{3+}$, $M_S$\,=\,+1) and electron (Cu$^{1+}$, $M_S$\,=\,-1) centers, the Cu$^{2+}$ ($M_S$\,=\,0)  center corresponds to a quantum magnetic center with conventional spin $s$\,=\,1/2.
 
The Hamiltonian of the model includes the on-site and inter-site density-density correlations ($\Delta$ and $V$), and the Ising spin exchange coupling ($J$):
\begin{equation}
\mathcal {H} = 
\Delta \sum_i^{\phantom{N}} S_{iz}^2 
+ V \sum_{\left\langle ij\right\rangle} S_{iz} S_{jz} 
+ J \sum_{\left\langle ij\right\rangle} P_{i0} s_{iz} s_{jz} P_{j0},
\label{H}
\end{equation}
where $P_{i0} = 1 - S^2_{iz}$ is a projection operator, which picks out the spin-magnetic Cu$^{2+}$ state, the sums run over sites of a two-dimensional square lattice, $\langle ij \rangle$ means the nearest neighbours. We assume the total charge constraint:
\begin{equation}
n = \frac{1}{N} \sum_i S_{iz} = const,
\end{equation} where $n$ is the density of the doped charged impurities in the system. Nonmagnetic charge impurities might serve as an annealed disorder and also affect on the critical behavior.

The paper is devoted to the study of the critical properties of this model for different values of parameters $\Delta$ and $n$. Recently~\cite{diffVJ} we have shown that the ground state phase diagram of the model strongly depends on the ratio $V/J$. Hereafter in this paper we focused on the ``strong'' exchange limit, when $V < J/4$.
\section{Methods}
The critical properties of the system with Hamiltonian (\ref{H}) have been analyzed by means of modified parallel Monte Carlo algorithm~\cite{PAVT} under periodic boundary conditions. The critical temperature $T_c$ of the phase transition to an ordered state was determined by intersection of the Binder cumulants $U_L = 1- \langle \mathcal{O}^4 \rangle_L / 3\langle \mathcal{O}^2 \rangle^2_L$ for different lattice sizes $L$~\cite{Binder}, where $\mathcal{O}$ is an antiferromagnetic or charge order parameter.

In order to estimate the value of the critical exponent for the specific heat $\alpha$ we made use of a data fitting to a power law $C \sim \vert (T-T_c)/T_c \vert^{-\alpha}$ near $T_c$ (the direct method). In addition, the finite-size scaling theory~\cite{Fisher} was used in the first order:
\begin{align}
C (T_c,L) &= C_0 L^{\alpha/\nu}, \label{powerscale}\\
C (T_c,L) &= C_0 \ln L \hspace{1cm}\text{for Ising critical behavior,}
\label{logscale}
\end{align}
where $C_0$ is constant, $\nu$ is the critical exponent for the correlation length, which can be obtained from expressions
\begin{equation}
V_n =  \frac{\langle \mathcal{O}^n E \rangle}{\langle \mathcal{O}^n \rangle} - \langle E \rangle,\quad
V_n = L^{1/\nu} g_{V_n},
\label{Vn}
\end{equation}
where $V_n$ is the logarithmic derivative of the order parameter $\mathcal{O}^n$, $g_{V_n}$ is constant~\cite{Vn}.
\section{Numerical Results}
\begin{figure}[h]
\centering
	\includegraphics[width=\linewidth]{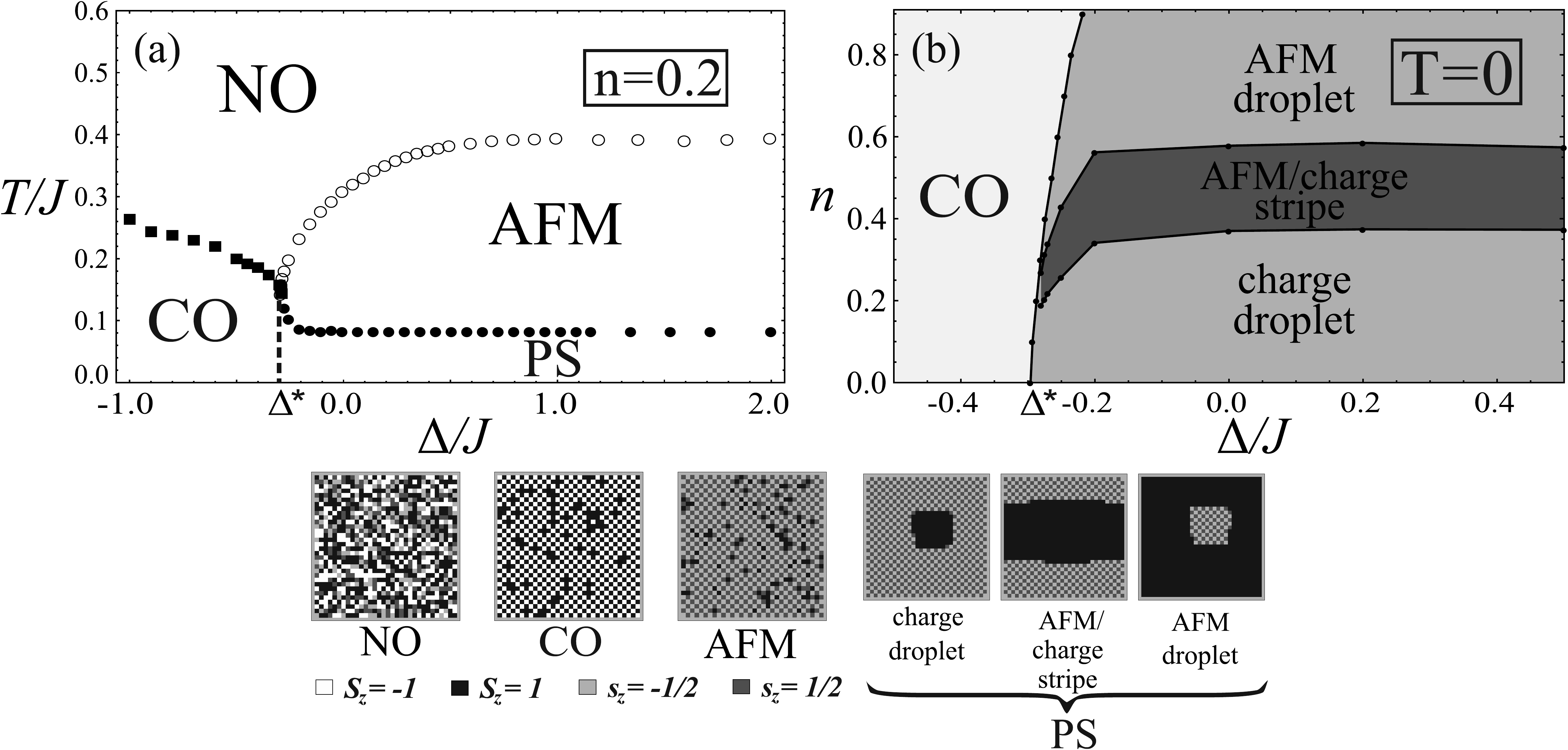}
	\caption{(a) The $T/J-\Delta$ phase diagram for $n=0.2$. (b) The ground state phase diagram. The snapshots of real states are made on a square lattice $32 \times 32$. NO denotes the non-ordered phase, CO - charge order, AFM - antiferromagnetic order, PS - phase separation}
	\label{PD}
\end{figure}

Fig.~\ref{PD} (left) shows the $T-\Delta/J$ phase diagram obtained at $n=0.2$. In case of $\Delta<\Delta^* = -0.3J$, a transition to the charge ordered (CO) phase occurs, while for $\Delta>\Delta^*$ the antiferromagnetic spin ordering (AFM) is formed as the temperature is lowered.  At the point $\Delta \approx \Delta^*$ both magnetic and charge subsystems have the same energy, so the ground state shows a degeneracy. The system evolves from the non-ordered paramagnetic phase (NO) to frustrated state with macroscopic phase separation on AFM and CO types of orderings. A complete ordering into an uniform state for $n=0$ reaches only at $T=0$, which indicates the quantum critical point. For $n \neq 0$ one can observe two sequential phase transitions for $\Delta>\Delta^*$: the first, AFM ordering diluted by randomly distributed charge impurities, and the second, unconventional phase separation (PS) described in~\cite{PS}. The configuration of the PS phase depends on the doping concentration $n$, as it is shown on the ground state phase diagram (Fig.~\ref{PD}, right). Depending on $n$, either a charge ``droplet'' or ``stripe'' is formed in the ground state, or antiferromagnetic droplet or stripe. Such configurations become preferable due to minimization of surface energy in a finite-size system with periodic boundary conditions. One should note that frustration also affects the PS states: there are macroscopic parts of the CO phase inside the charge droplets and stripes for $\Delta \approx \Delta^*$.

\begin{figure}
	\includegraphics[width=\linewidth]{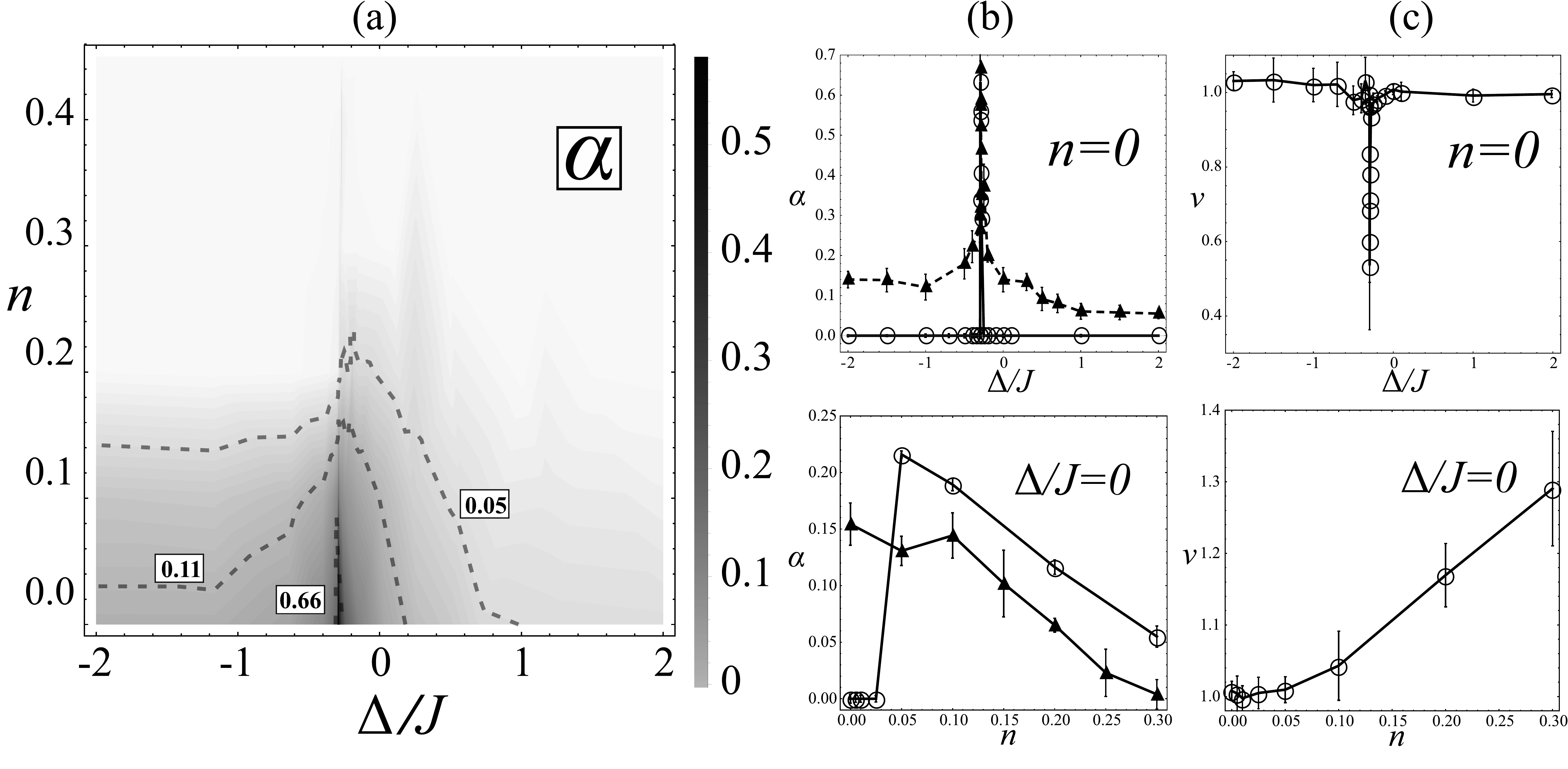}
	\caption{The critical exponents $\alpha$ and $\nu$ as functions of $n$ and $\Delta$. (a) The direct method results for $\alpha$. (b), (c) The dependencies of $\alpha$ and $\nu$ respectively on $\Delta$ (top) and $n$ (bottom). White circles correspond to the finite-size method, black triangles correspond to the direct method}
	\label{a}
\end{figure}

The dependencies of the critical exponents $\alpha$ and $\nu$ on $n$ and $\Delta$ are shown in Fig.\ref{a}. The values of $\alpha$ in contour plot and black triangles were determined by the direct method, whereas white circles correspond to the finite-size scaling results. Qualitatively both methods give similar results for $\alpha$. The values of $\nu$ presented in Fig.~\ref{a},c  were used to determine $\alpha$ in the finite-size scaling method. For $\vert \Delta \vert \gg \Delta^*$ and small $n$ the specific heat at $T_c$ grows slower than any positive degree, and we used expression (\ref{logscale}), which corresponds to $\alpha = 0$. It is worth noting that the similar method is used in~\cite{Landau}. Moreover, the values of $\nu$ in that case are equal to $1.0(0)$, which points to the 2D Ising universality class.
As one approaches the quantum critical point $\Delta^*$, the critical exponents dramatically change, and crossover between two types of critical behavior takes place. The critical exponents reach its extreme values close to the frustration point $\Delta \approx \Delta^*$. The bottom panels in Fig~\ref{a}, b and c show the dependencies of $\alpha$ and $\nu$ on $n$ respectively for $\Delta=0$. It can be seen that doping the impurities also leads to the crossover between the 2D Ising class and non-universal critical behavior. A similar crossover was suggested in review~\cite{Dotsenko}.
\section{Conclusions}
We presented a Monte Carlo study of a static 2D spin-pseudospin model that generalizes a 2D site-diluted antiferromagnetic Ising model with frustration due to competition between charge and spin degrees of freedom. An analysis of the ground state revealed various types of phase separation depending on the concentration of doped charge impurities $n$. The critical exponents for the specific heat $\alpha$ and for the correlation length $\nu$ have been determined. The high accuracy Monte Carlo simulations  allowed us to estimate $\alpha$ directly from the temperature dependence of the specific heat for a wide range of parameters $\Delta$ and $n$. Also for $n=0$ and $\Delta = 0$ the exponents are calculated in the framework of the finite-size scaling theory. Both methods give qualitatively similar dependencies of the critical exponents on the parameters of the system.

The model exhibits a crossover of the critical behavior from the 2D Ising universality class to non-universality. As one approaches a quantum critical point $\Delta^*$ as well as the doped charge density $n$ increases, the critical exponents begin to depend on the model parameters. The critical exponents reach its extreme values in a quantum critical point, and vary monotonically depending on doping.
\section{Acknowledgement}
The research was supported by the Government of the Russian Federation, Program $02.A03.21.0006$ and by the Ministry of Education and Science of the Russian Federation, projects N. $2277$ and $5719$ by RFBR according to the research project N. $18-32-00837/18$, scholarship of the president of the Russian Federation N. SP-$2278.2019.1$.
%


\end{document}